\documentclass{article}
\usepackage{spconf}
\usepackage{mathtools, amsmath,graphicx}
\usepackage{bm, bbm}
\usepackage{amssymb,amsfonts,amsthm}
\usepackage{siunitx}
\usepackage[hyphens]{url}
\usepackage{standalone,tikz}
\usetikzlibrary{patterns,shapes,arrows,calc,fit, decorations.pathreplacing}
\usetikzlibrary{backgrounds}
\usepackage{xcolor} %
\definecolor{RTX_neutral_medium}{RGB}{222,218,205}
\definecolor{RTX_neutral_light}{RGB}{237,236,231}
\definecolor{RTX_neutral_ultralight}{RGB}{251,251,251}
\usepackage{tikzscale}
\usepackage{textcomp}
\usepackage{enumerate}
\usepackage{cite}
\usepackage{siunitx}
\usepackage{booktabs,dcolumn,multirow,makecell,colortbl}
\newcommand\copyrighttext{%
\footnotesize \textcopyright 2023 IEEE. Personal use of this material is permitted. Permission from IEEE must be obtained for all other uses, in any current or future media, including reprinting/republishing this material for advertising or promotional purposes, creating new collective works, for resale or redistribution to servers or lists, or reuse of any copyrighted component of this work in other works.}
\renewcommand\copyrightnotice{%
\begin{tikzpicture}[remember picture,overlay]
\node[anchor=south,yshift=10pt] at (current page.south) {\fbox{\parbox{\dimexpr\textwidth-\fboxsep-\fboxrule\relax}{\copyrighttext}}};
\end{tikzpicture}%
}

\newtheorem{lemma}{Lemma}

\DeclareMathOperator*{\argmin}{arg\,min\,}
\title{JOINT MINIMUM PROCESSING BEAMFORMING AND NEAR-END LISTENING ENHANCEMENT}
\name{\begin{tabular}{c}Andreas J. Fuglsig$^{\star \dagger}$,
  Jesper Jensen$^{\dagger}$, 
  Zheng-Hua Tan$^{\dagger}$,
  Lars S. Bertelsen$^{\star}$,\\
  Jens Christian Lindof$^{\star}$,
  Jan Østergaard$^{\dagger}$, 
    \thanks{This work is partly supported by Innovation Fund Denmark Case no. 9065-00204B.}\end{tabular}}

\address{$^{\star}$ RTX A/S, Nørresundby, Denmark\\ 
$^{\dagger}$Aalborg University, Aalborg, Denmark}
\begin{document}
\ninept
\maketitle
\copyrightnotice
\begin{abstract}
We consider speech enhancement for signals picked up in one noisy environment that must be rendered to a listener in another noisy environment. For both far-end noise reduction and near-end listening enhancement, it has been shown that excessive focus on noise suppression or intelligibility maximization may lead to excessive speech distortions and quality degradations in favorable noise conditions, where intelligibility is already at ceiling level. Recently~\cite{zahedi_minimum_2021,fuglsig_minimum_2023} propose to remedy this with a minimum processing framework that either reduces noise or enhances listening a minimum amount given that a certain intelligibility criterion is still satisfied Additionally, it has been shown that joint consideration of both environments improves speech enhancement performance. In this paper, we formulate a joint far- and near-end minimum processing framework, that improves intelligibility while limiting speech distortions in favorable noise conditions. We  provide closed-form solutions to specific boundary scenarios and investigate performance for the general case using numerical optimization. We also show concatenating existing minimum processing far- and near-end enhancement methods preserves the effects of the initial methods. Results show that the joint optimization can further improve performance compared to the concatenated approach.
  

\end{abstract}
\section{Introduction}
\label{sec:intro}
Speech communication systems, including, e.g., mobile phones, hearing aids, and intercom systems, need to work in a variety of often noisy situations which can degrade intelligibility and quality.

\begin{figure}[!t]
  \centering
  \includegraphics[width=\columnwidth]{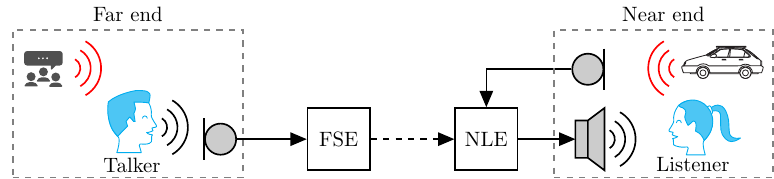}
  \caption{Basic communication system with Far-end Speech Enhancement (FSE) and  Near-end Listening Enhancement(NLE).\vspace*{-5mm}}
  \label{fig:system_concept}
\end{figure}

In speech communication systems, we may consider two distinct environments, cf. Fig.~\ref{fig:system_concept}: The far-end (the target talker location) and the near-end (the listener's location). 
Both environments may be susceptible to noise affecting the Speech Quality (SQ) and Intelligibility (SI) 
for the listener.
To counter this,
speech enhancement can be applied at both 
ends. Far-end Speech Enhancement (FSE) may employ single or multiple microphone noise reduction methods~\cite{zahedi_minimum_2021, doclo_acoustic_2010,loizou_speech_2013,gannot_consolidated_2017}. Near-end Listening Enhancement (NLE)~\cite{fuglsig_minimum_2023,kleijn_optimizing_2015,cooke_listening_2014}
 leverage knowledge of the near-end noise to 
 pre-process the received FSE signal 
for an optimal presentation with enhanced SI in the near-end background noise.
We note that headphone listening can utilize adaptive noise control (ANC) methods~\cite{george_advances_2013}. However, ANC with classic adaptive filtering falls short outside headphone use~\cite{li_near-end_2019}.
Thus, ANC is beyond our scope as we address speech presentation in an open environment.

NLE algorithms have conventionally aimed to solely enhance SI,
which may be beneficial at low SNRs but 
might diminish SQ at high SNRs due to excessive processing\cite{rennies_evaluation_2018,pricken_quality_2017,tang_study_2018,cooke_listening_2014}. 
Furthermore, many FSE methods are designed with a rationale
targeting eliminating all background noise to retain only clean speech,
causing potential excessive speech distortion or loss of contextual noise
~\cite{zahedi_minimum_2021}. 
Therefore, the works of \cite{zahedi_minimum_2021} and \cite{fuglsig_minimum_2023} apply a \emph{minimum processing principle} to FSE and NLE, respectively, where 
the noisy signal~\cite{zahedi_minimum_2021} or the signal received from the far-end~\cite{fuglsig_minimum_2023} is modified as little as possible while obtaining a desired level of SI. 
However, so far, the minimum processing principle has not been applied to situations where noise is present in both far-end and near-end environments simultaneously.
In fact, until recently,
 addressing disturbances in both the far-end and near-end settings was approached as separate tasks\cite{loizou_speech_2013,gannot_consolidated_2017,kleijn_optimizing_2015}.
However, recent work in \cite{niermann_joint_2017,fuglsig_joint_2022,khademi_intelligibility_2017,li_joint_2023,shifas_end--end_2022,zorila_quality_2017} have shown that optimizing SI by jointly addressing the noise in both environments is more effective than handling them as separate disjoint problems.

In this paper, we formulate a joint far- and near-end minimum processing framework,
which contrary to existing joint works only modifies the signal the minimum amount required to achieve a desired level of SI, and preserves SQ in favorable noise condition. Furthermore, it expands upon the existing minimum processing frameworks \cite{zahedi_minimum_2021,fuglsig_minimum_2023} by jointly considering the effects of FSE and NLE for both far- and near-end noise simultaneously.
Following \cite{zahedi_minimum_2021,fuglsig_minimum_2023} we minimize a mean-square error (MSE) processing penalty subject to an estimated SI constraint in terms of the Approximated Speech Intelligibility Index (ASII)~\cite{taal_optimal_2013}. We derive closed-form solutions for interesting special cases of the problem, and solve the general case using numerical optimization. We perform an experimental evaluation where we compare the proposed approach to a concatenation of minimum processing FSE~\cite{zahedi_minimum_2021} and minimum processing NLE~\cite{fuglsig_minimum_2023}.
The results show, that the proposed method is able to greatly improve SI up until a desired level in noisy conditions while also limiting speech distortions in favorable noise conditions. In addition, we see that concatenation preserves the minimum processing abilities of the individual methods while being able to improve both SI and SQ in various noise conditions. Finally, we show the joint approach is able to further improve performance compared to the concatenation.


\section{Signal Model}
\begin{figure}[!t]
  \centering
  \includegraphics[width=\columnwidth]{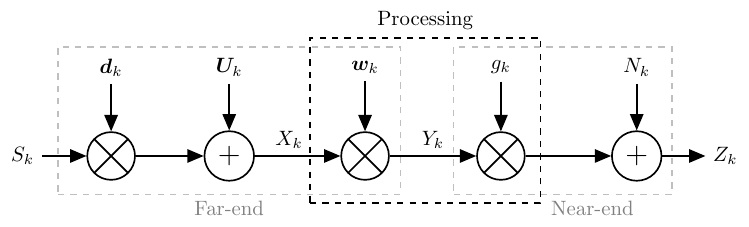}
  \caption{Signal model.\vspace*{-10mm}}
  \label{fig:signal_model}
\end{figure}
We consider a time-frequency domain representation of speech and noise signals with frequency index $k$. 
Since the statistics of the signals can be estimated online, and the mathematical framework can be applied on a per time-frame basis, we disregard the time index and assume we are considering a particular time frame, unless otherwise is stated.
The signal model in frequency bin $k$, cf. Fig.~\ref{fig:signal_model}, is given by
\begin{align}
  \bm{X}_k =\bm{d}_{k} S_{k} + \bm{U}_{k},~~
  Y_k =  \bm{w}_{k}^H\bm{X}_k, ~~
    Z_{k} = g_{k} Y_{k} + N_{k},
\end{align}
where $\bm{X}_{k}\in \mathbb{C}^M$ is the noisy multi-microphone signal, $S_{k}$ the clean speech signal recorded at the source location, $\bm{d}_k \in \mathbb{C}^M$ are acoustic transfer functions from the source to the microphones, and $\bm{U}_{k}\in \mathbb{C}^M$ is additive far-end noise, and $M$ is the number of microphones.
To increase SI and SQ, the noisy signal, $\bm{X}_{k}$, is linearly and spatially enhanced via a FSE noise reducing beamformer, $\bm{w}_k\in \mathbb{C}^M$, producing the modified signal $Y_{k}$. To further increase SI and SQ a NLE gain, $g_k\in \mathbb{R}_+$, is applied prior to playout. Finally, the signal,  $Z_{k}$, received at the near-end, is contaminated by ambient noise, $N_{k}$, in the environment. We assume the speech and noise processes are uncorrelated and zero-mean random processes, which are independent across frequency\cite{loizou_speech_2013}. We then have the speech distortion weighted covariance matrix $C_{X_k}^{(\mu)}$ of $\bm{X}_k$ is \cite{zahedi_minimum_2021},
\begin{align}
  C_{X_k}^{(\mu)}\triangleq C_{S_k}+\mu C_{U_k}=\sigma_{S_k}^2\bm{d}_k\bm{d}_k^H + \mu C_{U_k},
\end{align}
where $\sigma_{S_{k}}^2$ is the clean speech power spectrum level in time-frequency bin $k$ and $C_{U_k}$ is the far-end noise covariance matrix of $\bm{U}_k$ 
and where $\mu \in \mathbb{R}_+$ is the speech distortion weight\cite{brandstein_microphone_2001,doclo_acoustic_2010}.

In this paper, we process signals in perceptually relevant critical bands~\cite{american_national_standards_institute_methods_2017}, 
 with an individual non-negative filter weight, $\omega_{j,k}$, for each frequency bin-subband pair, where subbands are indexed by $j$ and frequencies with index $k$. We let 
 $\mathbb{B}_j$ denote the set of frequencies, $k$, that contribute to the $j$'th subband.


\section{Minimum Processing Concept}
To increase SI and SQ, the aim is to jointly determine a FSE beamformer, $\bm{w}_k$, for far-end noise reduction and a NLE gain, $g_k$, for pre-processing the signal before playout in near-end background noise.


Assume, as in \cite{zahedi_minimum_2021}, we are given a target reference signal, $S_k^{R}$, which may be the output of a beamformer with some desired properties, e.g., low speech distortion. Then 
for a particular subband, $j$, 
stack all $S_k^R$, $S_k$ and $Z_k$ for $k\in \mathbb{B}_j$
into vectors $\bm{S}_j^R, \bm{S}_j$ and $\bm{Z}_j$
\cite{zahedi_minimum_2021, fuglsig_minimum_2023}. 
Additionally, let $\mathcal{D}_j\left(\bm{S}_j^R, \bm{Z}_j\right)$ be a non-negative distortion measure (processing penalty) between the target reference signal, $S^R_k$, and the signal presented to the near-end listener, $Z_k$, and let $\mathcal{I}_j\left(\bm{S}_j^R, \bm{Z}_j\right)$ be a finite non-negative SI estimator of NLE-processed speech, $\bm{Z}_j$ in subband $j$.
Then, the joint far- and near-end minimum processing beamformer, $\bm{w}_k^{MP}$, and NLE gain, $g_k^{MP}$, in subband $j$ are defined as the solution to the following optimization problem:
\begin{equation}
	\begin{array}{lll}
		\displaystyle \argmin_{\{\bm{w}_k\}, \{g_k\}, k \in\mathbb{B}_j} 
    &\mathcal{D}_j\left(\bm{S}_j^R, \bm{Z}_j\right) & \mbox{s.t.}\quad \mathcal{I}_j\left(\bm{S}_j, \bm{Z}_j\right)\geq I'_j. 
	\end{array}\label{eq:gen_minproc_problem}
\end{equation}

Here we consider the combined effects of all noise sources with far-end noise reduction and near-end listening enhancement simultaneously. This is contrary to \cite{zahedi_minimum_2021} that only considers far-end noise reduction, and in a similar manner \cite{fuglsig_minimum_2023} that is only concerned with near-end listening enhancement under the assumptions of a clean far-end. Thus, instead of taking a classic blind concatenated approach, where we solve the two versions of \eqref{eq:gen_minproc_problem} proposed in \cite{zahedi_minimum_2021} and \cite{fuglsig_minimum_2023} in succession, while they are unaware of each other and the processing they apply. In our proposed joint approach, we solve \eqref{eq:gen_minproc_problem} directly, such that all noise sources and processing steps of $\bm{w}_k$ and $g_k$ are jointly taken into account at the same time.

\section{Joint Minimum Processing}
To avoid comb filtering effects and inspired by the results of \cite{zahedi_minimum_2021} and \cite{niermann_joint_2017}, we propose the following parameterized multichannel noise reduction vector (beamformer), that is fixed across an entire subband,
\begin{equation}
    \bm{w}_{j,k} \triangleq \alpha_j \bm{w}_k^{\mu_R}
    + (1-\alpha_j)\bm{w}_k^{\mu_0},\label{eq:def_beamformer_combine}
\end{equation}
as a solution to \eqref{eq:gen_minproc_problem}. Here the parameter $\alpha_j \in [0,1]$, and $\bm{w}_k^{\mu_R}$ and $\bm{w}_k^{\mu_0}$
are speech distortion weighted Multichannel Wiener Filters (MWFs)~\cite{doclo_acoustic_2010,zahedi_minimum_2021}  
\begin{equation}
	\bm{w}_k^{\mu} \triangleq \left( C_{X_k}^{(\mu)}\right)^{-1} \sigma_{S_k}^2\bm{d}_k,
\end{equation}
with pre-selected speech distortion weights, $\mu_R$ and $\mu_0$, such that the reference beamformer $\bm{w}_k^{\mu_R}$ has low speech distortion and $\bm{w}_k^{\mu_0}$ has high noise reduction~\cite{zahedi_minimum_2021}. 


Similarly to avoid comb filtering by the NLE gains, $g_k$, 
we assume they are fixed across an entire subband, i.e.,
\begin{align}
    g_k=g_i, \quad \forall k,i \in \mathbb{B}_j.
\end{align}
This is also in line with results of existing NLE literature~\cite{khademi_intelligibility_2017,niermann_joint_2017, fuglsig_joint_2022, kleijn_optimizing_2015}.
%

\subsection{Processing Penalty}
For the processing penalty, $\mathcal{D}_j(\cdot)$ we consider an MSE criterion~\cite{zahedi_minimum_2021,fuglsig_minimum_2023}. 
Since we want to have low speech distortion, we consider the reference signal, $S_k^R$, to be the output of the reference MWF, $\bm{w}_k^{\mu_R}$, which was chosen above to have the property of low distortion. 
Therefore, the minimum processing solution to \eqref{eq:gen_minproc_problem}, i.e., $\bm{w}_{j,k}$ and $g_k$, should minimize the distance to $\bm{w}_k^{\mu_R}$.
That is, the processing penalty must punish excessive difference to the reference signal caused by both the beamforming and NLE post gain. 
We note, that an obvious way to increase the near-end output SNR is to increase $g_k$ to infinity. However, this would lead to excessive speech distortions, infinite playback volume, and most importantly increase the difference to the reference signal leading to a violation of the minimum processing concept. 
Hence, we propose the following processing penalty,
\begin{align}
  \mathcal{D}_j(\bm{S}_j^{R}, \bm{Z}_j)
  &=\left(1-\alpha_j\right)^2+ \left(1-g_j\right)^2.\label{eq:total_penal}
\end{align}
Here the first term is the processing penalty incurred by the beamformer and pushes $\bm{w}_{j,k}$ close to $\bm{w}_k^{\mu_R}$. The second term is the penalty incurred by the NLE gain and pushes $g_k \bm{w}_{j,k}$ close to $\bm{w}_k^{\mu_R}$ and limits any speech distortions and excessive playback volume caused by the NLE gain.

\subsection{Performance Criteria}

We consider two different performance criteria; an intelligibility performance criterion and a new noise power criterion. 

\subsubsection{Intelligibility criterion}
We consider a performance criterion based on the ASII~\cite{taal_optimal_2013} as in \cite{fuglsig_minimum_2023} whereas \cite{zahedi_minimum_2021} uses SII. Letting $I_j$ be a given minimum requirement on the ASII subband SI performance~\cite{taal_optimal_2013}, 
the SI constraint in terms of the subband SNR, $\xi_j$, is~\cite[App.~C]{fuglsig_minimum_2023}
\begin{align}
  \xi_j \triangleq\tfrac{g_j^2 \delta_{S_j}(\alpha_j)}{g_j^2 \delta_{U_j}(\alpha_j) + \sigma_{\mathcal{N}_j}^2}, 
  \quad \xi_j \geq \frac{I_j}{1-I_j} \triangleq I_j^\xi,  
  \label{eq:def_I_xi}
\end{align}
where we consider $I_j^\xi \triangleq \frac{I_j}{1-I_j}$ as a target SNR, and $\delta_{S_j}(\alpha_j)$ and $\delta_{U_j}(\alpha_j)$ denote the processed speech and far-end noise power within one subband, $j$, for a given $\alpha_j$, respectively. By evaluating $\bm{w}_{j,k}^H C_{U_k}\bm{w}_{j,k}$ and filtering into subbands we have
\begin{flalign}
    \delta_{U_j}(\alpha_j)
   &= \alpha_j^2 \delta_{U_j}^{\mu_R} + (1-\alpha_j)^2 \delta_{U_j}^{\mu_0}+\alpha_j(1- \alpha_j) \delta_{U_j}^{cross},
   \label{eq:def_delta_U_alpha}
   \\
   \delta_{U_j}^{\mu}&\triangleq\textstyle\sum_{k\in \mathbb{B}_j}\omega_{j,k} \left(\bm{w}_k^{\mu}\right)^H C_{U_k}\bm{w}_k^{\mu},\\
   \delta_{U_j}^{cross}&\triangleq\textstyle\sum_{k\in \mathbb{B}_j}\omega_{j,k} 2\Re\left\{\left(\bm{w}_k^{\mu_0}\right)^H C_{U_k}\bm{w}_k^{\mu_R}\right\}.
\end{flalign}
%
%
A similar definition applies to the processed speech subband power, $\delta_{S_j}(\alpha_j)$. 
Since we modify the speech to increase SI, the subband SNR, $\xi_j$, is defined as the ratio of the \textit{processed} speech subband power to the total \textit{processed} noise power at the near-end listener~\cite{taal_optimal_2013,fuglsig_minimum_2023,american_national_standards_institute_methods_2017}. This is different to \cite{zahedi_minimum_2021}, where SNR is clean speech power relative to the MSE between $S$ and $Y$, i.e, all processing to the original speech is considered as a noise term and does not include near-end noise, $N$. In this work, the far-end SNR is defined as $ \delta_{S_j}(\alpha_j)/\delta_{U_j}(\alpha_j)$.
%
%
Now by defining the polynomial $p_{FSE}(\alpha_j)$, representing FSE SNR performance, as
\begin{flalign}
  p_{FSE}(\alpha_j)&\triangleq \delta_{S_j}(\alpha_j) - \delta_{U_j}(\alpha_j)  I_j^\xi&&\\
	&=\alpha_j^2 D_j^{\mu_R}  + (1-\alpha_j)^2 D_j^{\mu_0} + \alpha_j(1-\alpha_j)D_j^{cross}
  \hspace*{-20pt}
  &&\label{eq:f_con_def}
\end{flalign}
where $D_j^{\mu} \triangleq \delta_{S_j}^{\mu}-\delta_{U_j}^{\mu}I_j^\xi$, and $D_j^{cross} \triangleq \delta_{S_j}^{cross}-\delta_{U_j}^{cross}I_j^\xi$.
We can then write the constraint as
 \begin{align}
   \mathcal{I}_j = g_j^2 p_{FSE}(\alpha_j) &\geq \sigma_{\mathcal{N}_j}^2 I_j^\xi.\label{eq:full_audi_constraint}
   \end{align}



\subsubsection{Noise power criterion}
Since we consider far-end and near-end noise jointly, we have more knowledge about the processing and noise situation than in \cite{zahedi_minimum_2021} and \cite{fuglsig_minimum_2023}. Therefore, looking at \eqref{eq:def_I_xi}, we see that to increase the SNR and satisfy the audibility constraint, the processed far-end noise might need to overpower the near-end noise. However, depending on the noise powers this increase in SI may come at an undesired loss in SQ due to increased total noise levels. Therefore, to limit distortions caused by excessive noise levels, in the new joint approach we impose a constraint on the processed far-end noise power,
\begin{equation}
	10 \log_{10}\left(g_j^2 \delta_{U_j}(\alpha_j)\right) \leq 10 \log_{10} \sigma_{\mathcal{N}_j}^2 + \Delta_{U_j},\label{eq:noise_lim_con_logdomain}
\end{equation}
where the parameter $\Delta_{U_j}$ controls how many dB the processed far-end noise
can deviate from the near-end noise in subband, $j$. 


\subsection{Optimization Problem and Boundary Solutions}
From the above derivations we have that the joint far- and near-end minimum processing speech enhancement problem \eqref{eq:gen_minproc_problem} with the MSE processing penalty \eqref{eq:total_penal}, ASII performance constraint \eqref{eq:full_audi_constraint} and noise power constraint \eqref{eq:noise_lim_con_logdomain} is
	\begin{align}
		&\displaystyle \argmin_{{\alpha_j, g_j \in \mathbb{R}_+}}  
    \quad 
    \left(1-\alpha_j\right)^2+ \left(1-g_j\right)^2
    \tag{$P_0$}\label{eq:optimize_problem_theorem}
		\\
		\text{s.t.}~~
		& 
    \begin{aligned}[t]   
      &\mathcal{C}_1: g_j^2 p_{FSE}(\alpha_j) \geq \sigma_{\mathcal{N}_j}^2 I_j^\xi,\quad
      &\mathcal{C}_3: 0\leq \alpha_j \leq 1,
      \\
      &\mathcal{C}_2: g_j^2 \delta_{U_j}(\alpha_j) \leq \sigma_{\mathcal{N}_j}^2 10^{\Delta_{U_j}/10},\nonumber
      &\mathcal{C}_4: 1 \leq g_j. \nonumber
    \end{aligned}
	\end{align}

We can solve this optimization problem using a grid search algorithm.
Given the optimal solution $(\alpha_j^*, g_{j}^{MP})$, the optimum minimum processing beamformer is then given as
\begin{equation}
  \bm{w}_{j,k}^{MP} = \alpha_j^* \bm{w}_k^{\mu_R} + (1-\alpha_j^*)\bm{w}_k^{\mu_0}.
\end{equation}
%
From \eqref{eq:def_delta_U_alpha} and \eqref{eq:f_con_def},
we see that both the processed far-end noise power, $\delta_{U_j}$, 
and processed far-end SNR performance, $p_{FSE}$, 
include terms from each beamformer and a crossover term, and that the parameter $\alpha$ provides a trade off between the SNR/processing possible by the two candidate beamformers. Furthermore, we see that for $\alpha_j=0$ or $\alpha_j=1$, then $\bm{w}_{j,k} =\bm{w}_k^{\mu_R}$ or $\bm{w}_{j,k}=\bm{w}_k^{\mu_0}$, respectively, and thus the crossover terms vanish as well as the term accounting for the other beamformer. From this and inspection of the constraints we have the following lemma showing conditions for feasible boundary solutions. Proof omitted due to space limitations.


\begin{lemma}
  The beamformer combination weight, $\alpha_j^*=1$ is a solution to \eqref{eq:optimize_problem_theorem} under one of the two following conditions:
    (i) If ${D^{\mu_R}_j \geq \sigma_{\mathcal{N}_j}^2 I_j^\xi}$ and $\delta_{U_j}^{\mu_R}\leq \sigma_{\mathcal{N}_j}^2 10^{\Delta_{U_j}/10}$, with optimal NLE gain $g_j^*=1$.
    (ii) If $D^{\mu_R}_j \in(0,\sigma_{\mathcal{N}_j}^2 I_j^\xi)$ and $g_j^2 \delta_{U_j}^{\mu_R}\leq \sigma_{\mathcal{N}_j}^2 10^{\Delta_{U_j}/10}$, where the NLE gain is $g_j^2=\sigma_{\mathcal{N}_j}^2 I_j^\xi/D^{\mu_R}_j$.

  The beamformer combination weight, $\alpha_j^*=0$ is a solution to \eqref{eq:optimize_problem_theorem} under one of the two following conditions:
    (i) If $D^{\mu_0}_j \geq \sigma_{\mathcal{N}_j}^2 I_j^\xi$ and $\delta_{U_j}^{\mu_0}\leq \sigma_{\mathcal{N}_j}^2 10^{\Delta_{U_j}/10}$, with NLE gain $g_j^*=1$.
    (ii) If $D^{\mu_0}_j \in (0,\sigma_{\mathcal{N}_j}^2 I_j^\xi)$ and $g_j^2 \delta_{U_j}^{\mu_0}\leq \sigma_{\mathcal{N}_j}^2 10^{\Delta_{U_j}/10}$, where the NLE gain is $g_j^2=\sigma_{\mathcal{N}_j}^2 I_j^\xi/D^{\mu_0}_j$.
\end{lemma}

Depending on the subband definition, multiple frequencies may contribute to multiple subbands indexed by $\mathbb{F}_k$. Therefore, the optimum beamformer $\bm{w}_{j,k}^{MP}$ and NLE gain $ g_j^{MP}$ may also contribute to multiple subbands. Letting $\eta_{j,k}$ denote the weight that accounts for the impact of this contribution, the optimal frequency dependent beamformer and NLE gain, respectively, are
\begin{align}
  \bm{w}_{k}^{MP}=\sum_{j \in \mathbb{F}_k}\eta_{j,k}\bm{w}_{j,k}^{MP} \quad \text{and} \quad
 g_k^{MP}=\sum_{j \in \mathbb{F}_k}\eta_{j,k}g_{j}^{MP}.
\end{align}

Depending on the subband noise powers, the constraints of \eqref{eq:optimize_problem_theorem} may be infeasible. For example, $\mathcal{C}_1$ is infeasible if the far-end noise cannot be sufficiently reduced to produce a feasible far-end SNR. Similarly, $\mathcal{C}_2$ is infeasible if the remaining far-end noise power is too high compared to the near-end noise power.
We propose three ways to handle the infeasible situations.

If $\mathcal{C}_1$ is infeasible:
First find an $\alpha_j^*$ that maximizes the far-end SNR, ${\delta_{S_j}(\alpha)}/{\delta_{U_j}(\alpha)}$. Then to increase SI as function of $g_j$ for a fixed $\alpha_j^*$, select the NLE gain, $g_j^*$, such that the near-end noise does not decrease the SNR coming from the far-end more than ${\Delta_{N_j}>0}$\,dB. The gain, $g_j^*$ is then clipped according to $\mathcal{C}_2$ and $\mathcal{C}_4.$

If $\mathcal{C}_2$ is infeasible: Let $g_j^*=1$ and find an $\alpha_j^*$ so as the near-end SNR, $\xi_j$, is close to $I_j^\xi$ to approach satisfying both $\mathcal{C}_1$ and $\mathcal{C}_2$.

If the intersected constraints $\mathcal{C}_1 \cap \mathcal{C}_2$ are infeasible:
Find an $\alpha_j^*$ such that the processed near-end SNR is close to $I_j^\xi$ while adhering to $\mathcal{C}_2$. Then select the NLE gain, $g_j^*$, such that the processed near-end SNR is maximized while minimizing $g_j$ and satisfying $\mathcal{C}_2$.

\section{Experimental Evaluation}
We compare performance between the proposed joint minimum processing method and the concatenation of the FSE~\cite{zahedi_minimum_2021} and NLE~\cite{fuglsig_minimum_2023}.
We investigate two scenarios: (i) When the target talker is in a babble noise setting, e.g., office environment, and the listener is driving in a car, and (ii) the reverse scenario where the talker is in car noise and the listener is in babble noise.
The FSE beamformer of \cite{zahedi_minimum_2021} is also parameterized according to two $\mu$-MWF beamformers, and these beamformers are selected to be the same as in the proposed method, where we have $\mu_R=0,~\mu_0=5$. The per-band audibility target input, $I_j$, is weighted from a total SII target, $A^*=0.7$, using the band importance functions of the SII~\cite{american_national_standards_institute_methods_2017}, cf.~\cite[Sec. IV.B]{fuglsig_minimum_2023}, in both the proposed and reference method. Similarly, for the proposed method, the parameters, $\Delta_{U_j}$ and $\Delta_{N_j}$ are weighted for each subband from a single value of $\Delta_U$, and $\Delta_N$, respectively. Through informal listening tests we have selected $\Delta_U=\SI{12}{\dB}$, and $\Delta_N=\SI{10}{\dB}$ for scenario (i), and $\Delta_U=\SI{0}{\dB}$, and $\Delta_N=\SI{10}{\dB}$ for scenario (ii).

\subsection{Experimental Setup}
The far-end room dimensions are $3 \times 4 \times 3$\,\si{m^3}, with the target talker located at $[1.50, 3.00, 1]$\,\si{m}, 
and three noise sources located at $[0.50, 1.00, 1]$\,\si{m}, $[0.75, 3.00, 1]$\,\si{m} and $[3.00, 1.60, 1]$\,\si{m}.
The far-end has two microphones at $[1.50, 2.00, 1]$\,\si{m} and $[1.50, 2.02, 1]$\,\si{m}. Each microphone is also subject to a $\SI{60}{dB}$ SNR white noise. 
The time-frequency representations of the speech and noise signals are based on a DFT with $\SI{32}{ms}$ windows with $50\%$ overlap. We consider a total of $J = 30$ critical bands with center frequencies linearly spaced  on the equivalent rectangular bandwidth scale from \SIrange{150}{8000}{\Hz} derived according to \cite{van_de_par_perceptual_2005}.
For simplicity, signals are processed in a time-invariant manner and power spectrums are evaluated as the long-term average across time-frames.
The long-term power spectrums of the speech and noise are assumed to be known along with the room transfer functions, that are generated without reverberation using \cite{allen_image_1979}.
The speech material is sentences from the TIMIT~\cite{garofolo_timit_1993} test set sampled at $\SI{16}{kHz}$. Performance is evaluated across a total of $10$ trials, where for each trial a speaker is selected randomly without replacement, and a random sentence is selected for the given speaker. We then average the performance across the trials for each combination of noise, SNR and enhancement method.

\subsection{Results}
Estimated SI and SQ performance is measured with ESTOI~\cite{jensen_algorithm_2016} and PESQ~\cite{itu-t_recommendation_2001}. Table~\ref{tab:results} shows scores for the proposed and concatenated method alongside the unprocessed performance, with the best ESTOI and PESQ scores highlighted for each SNR and noise pair.

The results indicate that the proposed joint method and the concatenation method generally exhibit similar performance, as expected due to their overall similarity. 
However, in severe noise with low SNRs, where the unprocessed performance is very low, the proposed joint method overall outperforms the blind concatenation in ESTOI.
As the SNRs increase, the unprocessed SQ and SI score rise naturally. Here, when the noise situation is more favorable, both methods are able to utilize their minimum processing designs and limit distortions to better preserve the natural SQ and SI, as seen by how the ESTOI performance is close to the high unprocessed scores, while the PESQ scores still improve or stay close to the unprocessed scores. 
Since both the proposed joint method and the individual steps in the concatenation of \cite{zahedi_minimum_2021} and \cite{fuglsig_minimum_2023} are designed with minimum processing in mind, we did not expect a big difference in SQ performance at high SNRs. 

We also see, that the concatenation of \cite{zahedi_minimum_2021} and \cite{fuglsig_minimum_2023} preserves the effects of the individual methods, i.e., the signal is only processed the minimum required amount to obtain a desired SI at the far- and near-end respectively, and preserves SQ in favorable noise conditions.

For far-end car noise we observe, that the blind method is able to increase PESQ slightly more than the proposed method, we expect this is caused by the slight variations between the constraints in the two methods. Hence, further benefits might be gained from adjusting the proposed method accordingly.
However, because the concatenation is blind the FSE beamformer \cite{zahedi_minimum_2021} may not remove a sufficient amount of noise for the NLE in \cite{fuglsig_minimum_2023} to be able to achieve the desired SI. Similarly, because the NLE in \cite{fuglsig_minimum_2023} is blind to noise coming from the far-end it might not provide a sufficiently high gain as it mistakes noise for speech.
On the other hand, the proposed joint method can achieve a higher SI performance because it has access to all noise and processing information simultaneously.

\begin{table}[!t]
  \centering
  \renewcommand{\arraystretch}{1.12}
  \resizebox{\columnwidth}{!}{%
  \begin{tabular}{@{}ccrrc@{\hspace{2pt}}rrrc@{\hspace{2pt}}rrr@{}}
    \toprule
    \multicolumn{2}{l}{Noise}
    &\multicolumn{2}{c}{SNR}&&\multicolumn{3}{c}{ESTOI} &&\multicolumn{3}{c}{PESQ} \\
    \cmidrule(r){1-2}\cmidrule(lr){3-4} \cmidrule(lr){6-8} \cmidrule{10-12}
    FE & NE & FE & NE && \multicolumn{1}{c}{Prop.} & \multicolumn{1}{c}{Blind} & \multicolumn{1}{c}{Unp.} && \multicolumn{1}{c}{Prop.} & \multicolumn{1}{c}{Blind} & \multicolumn{1}{c}{Unp.}\\\midrule
    B& C&$\SI{0}{\dB}$ & $\SI{-25}{\dB}$ && $\bm{0.482}$& $0.419$ &$0.380$ && $\bm{1.052}$ & $1.030$ & $1.028$\\
    B& C&$\SI{0}{\dB}$ &   $\SI{0}{\dB}$ && ${0.569}$   & $0.512$ & $\bm{0.629}$ && $1.292$& $1.282$&$\bm{1.320}$\\
    B& C&$\SI{0}{\dB}$ &  $\SI{15}{\dB}$ && ${0.618}$   & $0.604$ & $\bm{0.680}$ && $1.400$& $\bm{1.465}$ & $1.389$\\[1.5mm]

    B& C&$\SI{-10}{\dB}$ & $\SI{-30}{\dB}$ && $\bm{0.262}$& $0.231$&$0.214$&& $\bm{1.044}$& $1.030$&$1.040$\\
    B& C&  $\SI{0}{\dB}$ & $\SI{-30}{\dB}$ && $\bm{0.478}$& $0.416$&$0.314$&& $\bm{1.050}$& $1.030$&$1.024$\\
    B& C& $\SI{10}{\dB}$ & $\SI{-30}{\dB}$ && $\bm{0.529}$& $0.506$&$0.343$&& $\bm{1.034}$& $1.031$&$1.024$\\

    \midrule

    C& B&$\SI{-10}{\dB}$ & $\SI{-20}{\dB}$ && $\bm{0.533}$& $0.500$&$0.036$&& $1.286$& $\bm{1.317}$&$1.078$\\
    C& B&$\SI{-10}{\dB}$ &   $\SI{0}{\dB}$ && $\bm{0.542}$& $0.511$&$0.460$&& $1.284$& $\bm{1.318}$&$1.147$\\
    C& B&$\SI{-10}{\dB}$ &  $\SI{20}{\dB}$ && $0.684$  & $0.679$&$\bm{0.748}$&& $1.832$ & $\bm{2.012}$&$1.542$\\[1.5mm]

    C& B&$\SI{-20}{\dB}$ & $\SI{-10}{\dB}$ && $\bm{0.470}$& $0.439$&$0.170$      && $1.254$& $\bm{1.291}$ &$1.109$\\
    C& B&  $\SI{0}{\dB}$ & $\SI{-10}{\dB}$ && $\bm{0.544}$& $0.535$&$0.194$      &&$\bm{1.326}$& $1.325$&$1.128$\\
    C& B& $\SI{20}{\dB}$ & $\SI{-10}{\dB}$ && $\bm{0.544}$ & $\bm{0.544}$&$0.195$&& $\bm{1.326}$& $\bm{1.326}$&$1.128$\\

     \bottomrule 
  \end{tabular}
  }
  \caption{ESTOI and PESQ scores for the proposed joint method and the concatenation of \cite{zahedi_minimum_2021} and \cite{fuglsig_minimum_2023} for various SNRs along with the unprocessed performance. Here B is babble noise and C is car noise.\vspace*{-5mm}} 
  \label{tab:results}
\end{table}

\section{Conclusion}
We formulated a joint far- and near-end minimum processing framework, where the beamformed and near-end listening enhanced output signal is optimized to have the minimum amount of processing artifacts with the constraint that an intelligibility performance criterion is satisfied. The proposed method adapts to environmental noise conditions and focuses on improving intelligibility in very noisy conditions, and, by the minimum processing concept, quality in favorable noise conditions. We show closed-form solutions to interesting special cases of the optimization problem.
Additionally, we show that speech enhancement using a blind concatenation of the existing far- and near-end minimum processing frameworks \cite{zahedi_minimum_2021} and \cite{fuglsig_minimum_2023} preserves the minimum processing abilities of the individual methods, and that the concatenation is also able to improve both intelligibility and quality in various noise conditions.
Results also show that the proposed joint method outperforms the simple blind concatenation in terms of intelligibility enhancement because the proposed method considers all noise sources and processing steps simultaneously.


\bibliographystyle{IEEEbib}
\bibliography{ICASSP_2024_bib}

\begin{thebibliography}{10}

\bibitem{zahedi_minimum_2021}
Adel Zahedi, Michael~S. Pedersen, Jan Østergaard, Thomas~U. Christiansen, Lars
  Bramsløw, and Jesper Jensen,
\newblock ``Minimum {Processing} {Beamforming},''
\newblock {\em IEEE/ACM Transactions on Audio, Speech, and Language
  Processing}, vol. 29, pp. 2710--2724, 2021.

\bibitem{fuglsig_minimum_2023}
Andreas~Jonas Fuglsig, Jesper Jensen, Zheng-Hua Tan, Lars~Søndergaard
  Bertelsen, Jens~Christian Lindof, and Jan Østergaard,
\newblock ``Minimum {Processing} {Near}-{End} {Listening} {Enhancement},''
\newblock {\em IEEE/ACM Transactions on Audio, Speech, and Language
  Processing}, vol. 31, pp. 2233--2245, 2023.

\bibitem{doclo_acoustic_2010}
Simon Doclo, Sharon Gannot, Marc Moonen, and Ann Spriet,
\newblock ``Acoustic {Beamforming} for {Hearing} {Aid} {Applications},''
\newblock in {\em Handbook on {Array} {Processing} and {Sensor} {Networks}},
  pp. 269--302. John Wiley \& Sons, Ltd, 2010.

\bibitem{loizou_speech_2013}
Philipos~C. Loizou,
\newblock {\em Speech {Enhancement}: {Theory} and {Practice}},
\newblock CRC Press, Boca Raton, FL, 2nd edition, 2013.

\bibitem{gannot_consolidated_2017}
Sharon Gannot, Emmanuel Vincent, Shmulik Markovich-Golan, and Alexey Ozerov,
\newblock ``A {Consolidated} {Perspective} on {Multimicrophone} {Speech}
  {Enhancement} and {Source} {Separation},''
\newblock {\em IEEE/ACM Transactions on Audio, Speech, and Language
  Processing}, vol. 25, no. 4, pp. 692--730, Apr. 2017.

\bibitem{kleijn_optimizing_2015}
W.~Bastiaan Kleijn, Joao~B. Crespo, R.~C. Hendriks, Petko~N. Petkov, Bastian
  Sauert, and Peter Vary,
\newblock ``Optimizing {Speech} {Intelligibility} in a {Noisy} {Environment}:
  {A} unified view,''
\newblock {\em IEEE Signal Processing Magazine}, vol. 32, no. 2, pp. 43--54,
  Mar. 2015.

\bibitem{cooke_listening_2014}
Martin Cooke, Simon King, Maëva Garnier, and Vincent Aubanel,
\newblock ``The listening talker: {A} review of human and algorithmic
  context-induced modifications of speech,''
\newblock {\em Computer Speech \& Language}, vol. 28, no. 2, pp. 543--571, Mar.
  2014.

\bibitem{george_advances_2013}
Nithin~V. George and Ganapati Panda,
\newblock ``Advances in active noise control: {A} survey, with emphasis on
  recent nonlinear techniques,''
\newblock {\em Signal Processing}, vol. 93, no. 2, pp. 363--377, Feb. 2013.

\bibitem{li_near-end_2019}
Gang Li, Ruimin Hu, Xiaochen Wang, and Rui Zhang,
\newblock ``A near-end listening enhancement system by {RNN}-based noise
  cancellation and speech modification,''
\newblock {\em Multimedia Tools and Applications}, vol. 78, no. 11, pp.
  15483--15505, June 2019.

\bibitem{rennies_evaluation_2018}
Jan Rennies, Arna Pusch, Hening Schepker, and Simon Doclo,
\newblock ``Evaluation of a near-end listening enhancement algorithm by
  combined speech intelligibility and listening effort measurements,''
\newblock {\em The Journal of the Acoustical Society of America}, vol. 144, no.
  4, pp. EL315--EL321, Oct. 2018.

\bibitem{pricken_quality_2017}
Robin Pricken, Marcel Wältermann, Eva Parotat, Michal Soloducha, and Alexander
  Raake,
\newblock ``Quality {Aspects} of {Near}-{End} {Listening} {Enhancement}
  {Approaches} in {Telecommunication} {Applications},''
\newblock in {\em Proceedings of {DAGA} 2017}, Kiel, 2017, pp. 872--875, German
  Acoustical Society (DEGA).

\bibitem{tang_study_2018}
Yan Tang, Christopher Arnold, and Trevor~J. Cox,
\newblock ``A {Study} on the {Relationship} between the {Intelligibility} and
  {Quality} of {Algorithmically}-{Modified} {Speech} for {Normal} {Hearing}
  {Listeners},''
\newblock {\em Journal of Otorhinolaryngology, Hearing and Balance Medicine},
  vol. 1, no. 1, pp. 10, June 2018.

\bibitem{niermann_joint_2017}
Markus Niermann, Peter Jax, and Peter Vary,
\newblock ``Joint {Near}-{End} {Listening} {Enhancement} and far-end noise
  reduction,''
\newblock in {\em 2017 {IEEE} {International} {Conference} on {Acoustics},
  {Speech} and {Signal} {Processing} ({ICASSP})}. Mar. 2017, pp. 4970--4974,
  IEEE.

\bibitem{fuglsig_joint_2022}
Andreas~Jonas Fuglsig, Jan Østergaard, Jesper Jensen, Lars~Søndergaard
  Bertelsen, Peter Mariager, and Zheng-Hua Tan,
\newblock ``Joint {Far}- and {Near}-{End} {Speech} {Intelligibility}
  {Enhancement} {Based} on the {Approximated} {Speech} {Intelligibility}
  {Index},''
\newblock in {\em {ICASSP} 2022 - 2022 {IEEE} {International} {Conference} on
  {Acoustics}, {Speech} and {Signal} {Processing} ({ICASSP})}, Singapore, May
  2022, pp. 7752--7756, IEEE.

\bibitem{khademi_intelligibility_2017}
Seyran Khademi, Richard~C. Hendriks, and W.~Bastiaan Kleijn,
\newblock ``Intelligibility {Enhancement} {Based} on {Mutual} {Information},''
\newblock {\em IEEE/ACM Transactions on Audio, Speech, and Language
  Processing}, vol. 25, no. 8, pp. 1694--1708, Aug. 2017.

\bibitem{li_joint_2023}
Haoyu Li, Yun Liu, and Junichi Yamagishi,
\newblock ``Joint {Noise} {Reduction} and {Listening} {Enhancement} for
  {Full}-{End} {Speech} {Enhancement},''
\newblock in {\em {ICASSP} 2023 - 2023 {IEEE} {International} {Conference} on
  {Acoustics}, {Speech} and {Signal} {Processing} ({ICASSP})}. June 2023, pp.
  1--5, IEEE.

\bibitem{shifas_end--end_2022}
Muhammed~P.V. Shifas, Cătălin Zorilă, and Yannis Stylianou,
\newblock ``End-to-{End} {Neural} {Based} {Modification} of {Noisy} {Speech}
  for {Speech}-in-{Noise} {Intelligibility} {Improvement},''
\newblock {\em IEEE/ACM Transactions on Audio, Speech, and Language
  Processing}, vol. 30, pp. 162--173, 2022.

\bibitem{zorila_quality_2017}
Tudor-Cătălin Zorilă and Yannis Stylianou,
\newblock ``On the {Quality} and {Intelligibility} of {Noisy} {Speech}
  {Processed} for {Near}-{End} {Listening} {Enhancement},''
\newblock in {\em Interspeech 2017}. Aug. 2017, pp. 2023--2027, ISCA.

\bibitem{taal_optimal_2013}
Cees~H. Taal, Jesper Jensen, and Arne Leijon,
\newblock ``On {Optimal} {Linear} {Filtering} of {Speech} for {Near}-{End}
  {Listening} {Enhancement},''
\newblock {\em IEEE Signal Processing Letters}, vol. 20, no. 3, pp. 225--228,
  Mar. 2013.

\bibitem{brandstein_microphone_2001}
Michael Brandstein and Darren Ward, Eds.,
\newblock {\em Microphone arrays: signal processing techniques and
  applications},
\newblock Digital signal processing. Springer, New York, 2001.

\bibitem{american_national_standards_institute_methods_2017}
{American National Standards Institute},
\newblock {\em Methods for {Calculation} of the {Speech} {Intelligibility}
  {Index}},
\newblock Acoustical Society of America, New York, N.Y, {ANSI} s.35-1997
  edition, 2017.

\bibitem{van_de_par_perceptual_2005}
Steven van~de Par, Armin Kohlrausch, Richard Heusdens, Jesper Jensen, and
  Søren~Holdt Jensen,
\newblock ``A {Perceptual} {Model} for {Sinusoidal} {Audio} {Coding} {Based} on
  {Spectral} {Integration},''
\newblock {\em EURASIP Journal on Advances in Signal Processing}, vol. 2005,
  no. 9, pp. 1292--1304, June 2005.

\bibitem{allen_image_1979}
Jont Allen and David Berkley,
\newblock ``Image method for efficiently simulating small-room acoustics,''
\newblock {\em The Journal of the Acoustical Society of America}, vol. 65, pp.
  943--950, Apr. 1979.

\bibitem{garofolo_timit_1993}
John~S. Garofolo, Lori~F. Lamel, William~M. Fisher, Jonathan~G. Fiscus,
  David~S. Pallett, Nancy~L. Dahlgren, and Victor Zue,
\newblock ``{TIMIT} {Acoustic}-phonetic {Continuous} {Speech} {Corpus},''
\newblock {\em Linguistic Data Consortium}, 1993.

\bibitem{jensen_algorithm_2016}
J.~Jensen and Cees~H. Taal,
\newblock ``An {Algorithm} for {Predicting} the {Intelligibility} of {Speech}
  {Masked} by {Modulated} {Noise} {Maskers},''
\newblock {\em IEEE/ACM Transactions on Audio, Speech, and Language
  Processing}, vol. 24, no. 11, pp. 2009--2022, Nov. 2016.

\bibitem{itu-t_recommendation_2001}
ITU-T,
\newblock ``Recommendation {P}.862: {Perceptual} evaluation of speech quality
  ({PESQ}),''
\newblock Recommendation ITU-T P.862, International Telecommunication Unition,
  Feb. 2001.

\end{thebibliography}

\end{document}